# Polar-Domain Volume as a Unified Descriptor of Transport in Ionic Liquids

Ganesh K. Rajahmundry and Tarak K. Patra*
Department of Chemical Engineering
Indian Institute of Technology Madras, Chennai, TN 600036, India

## Abstract

Developing molecular-scale structural descriptors that can quantitatively predict the macroscopic transport properties of ionic liquids (ILs) remains a longstanding question, owing to the complex interplay between molecular interactions, nanoscale organization, and collective ion dynamics. Here, we use all-atom molecular dynamics simulations to establish quantitative structure–property relationships between the equilibrium microstructure and two key transport properties—viscosity and ionic conductivity—in a series of imidazolium-based ILs with chemically distinct anions and varying alkyl-chain lengths. While viscosity and ionic conductivity exhibit distinct dependencies on alkyl-chain length and anion chemistry, these seemingly different trends largely collapse onto unified correlations when expressed in terms of the mean polar-domain volume. In particular, both transport properties exhibit systematic power-law scaling with the mean polar-domain volume fraction, revealing a common structural origin underlying the variations in ion and momentum transport across chemically distinct ILs. These results establish polar-domain volume as a physically motivated molecular-scale descriptor and provide a general structure–property framework for connecting nanoscale organization to macroscopic transport properties in ILs.

*Authors to correspond, E-mail: tpatra@iitm.ac.in

Ionic liquids (ILs) are organic salts with melting points below 100 °C, and those that remain liquid near room temperature are commonly referred to as room-temperature ionic liquids (RTILs).[1–4] Their negligible vapor pressure, high thermal and chemical stability, wide electrochemical windows, and tuneable physicochemical properties have made ILs attractive for a broad range of applications, including electrolytes for electrochemical systems, separations, catalysis, and other emerging technologies.[5,6] Among RTILs, imidazolium-based ILs are extensively studied[7,8] because their physicochemical properties can be systematically tuned through variations in cation structure, viz., alkyl-chain length, and anion chemistry. In particular, ionic conductivity and viscosity are highly sensitive to molecular architecture, making these systems useful model platforms for understanding the molecular origins of transport in complex ionic fluids.[9–11]

An important characteristic of imidazolium-based ILs is their intrinsic nanoscale structural heterogeneity. The amphiphilic nature of the constituent ions drives spatial segregation of the charged and nonpolar components, resulting in polar and nonpolar domains on the nanometer scale.[12,13] The extent and morphology of this segregation depend strongly on molecular structure.[14] For example, Padua and co-workers demonstrated the formation of nonpolar domains in 1-alkyl-3-methylimidazolium hexafluorophosphate ([$C_n$mim][$PF_6$]), with the nonpolar regions becoming increasingly connected as the alkyl-chain length increases.[15] A large alkyl tail can act as a separator of charge networks.[16] Experimental measurements and subsequent simulation studies have provided further evidence for nanoscale structural organization and heterogeneity in imidazolium-based ILs.[17,18] The physicochemical properties are intricately connected to the polar and nonpolar domain sizes of ILs.[19] These observations establish that ILs cannot always be viewed as structurally homogeneous ionic fluids; rather, their transport properties may emerge from a complex interplay between local molecular organization, domain morphology, and collective ion dynamics. Several studies have independently established that both the nanostructure and transport properties of imidazolium-based ILs depend strongly on alkyl-chain length and anion identity. Ionic conductivity generally decreases with increasing alkyl-chain length, whereas viscosity exhibits a strong dependence on both cation and anion chemistry. [20–24] In particular, chloride ion (Cl), hexafluorophosphate ion ($PF_6$), and bis(trifluoromethanesulfonyl)imide ion (TFSI)-based ILs can exhibit substantially different viscosities, reflecting differences in ion–ion interactions, coordination, and molecular packing. Although viscosity and ionic conductivity are intrinsically coupled through the dynamics of the liquid, it remains unclear whether their distinct molecular origins can be described by a common transport or structural parameter. Previous studies indicated that the ion pair lifetime can serve as a unified descriptor of ion conductivity for a large number of ILs.[25]

Building upon these previous works, we ask the question whether the complex molecular organisation of ILs can be reduced to a physically meaningful structural descriptor that quantitatively captures their transport behaviour. In particular, it remains unclear how the characteristic size and distribution of polar and nonpolar domains influence ion mobility and viscous relaxation, and whether these effects can be described using a common structural variable across chemically distinct ILs. This question is particularly important because changing the alkyl-chain length or anion identity simultaneously modifies several molecular-scale features, including ion coordination, intermolecular interactions, packing, and domain organization. Consequently, correlations based on a single chemical variable, such as alkyl chain length, may not capture a general structure–property relationship applicable across different IL chemistries. Developing a structural descriptor that incorporates these coupled molecular effects could therefore provide a more general framework for connecting nanoscale organization to macroscopic transport.

In this work, we use all-atom molecular dynamics (MD) simulations to establish quantitative relationships between the molecular microstructure and transport properties of imidazolium-based ILs. We investigate nine ILs comprising three cations—1-butyl-3-methylimidazolium (BMIM), 1-hexyl-3-methylimidazolium (HMIM), and 1-octyl-3-methylimidazolium (OMIM)—and three chemically distinct anions, Cl, $PF_6$, and TFSI. For each system, equilibrium MD trajectories are generated using a well-established molecular force field, from which we characterize molecular packing, ion coordination, nanoscale domain organization, viscosity, and ionic conductivity. We quantify the heterogeneous microstructure of each IL in terms of the distribution of polar-domain volumes. Rather than treating the individual chemical components as independent descriptors, we examine whether a characteristic measure of the polar-domain structure can unify the transport behavior across the different molecular architectures and anion chemistries. We find that both ionic conductivity and viscosity exhibit strong and systematic correlations with the mean polar-domain volume. Remarkably, despite the distinct dependencies of these transport properties on alkyl-chain length and anion identity, the data collapse onto common scaling relationships when expressed in terms of the mean polar-domain volume. The ionic conductivity increases with increasing polar-domain volume, whereas the viscosity decreases, with both properties exhibiting power-law scaling. These results demonstrate that the mean polar-domain volume provides a physically motivated and efficient microstructural descriptor for transport in imidazolium-based ILs. More broadly, our findings suggest that nanoscale domain organization provides a common structural basis for understanding both ion and momentum transport in chemically diverse ILs. The resulting structure–property framework offers a route toward connecting molecular-scale organization with macroscopic transport behavior and may provide a general strategy for identifying predictive structural descriptors in complex molecular fluids.

We use the OPLSAA[26] force field to model ILs. The pairwise interactions are shifted to ramp the energy smoothly to zero between 13 Å and 15 Å. Each IL system contains 125 cations and 125 anions. The size of the systems varies from 3250 atoms (BMIM-Cl) to 6500 atoms (OMIM-TFSI). We begin with a randomly generated configuration, and perform a 1ns simulation at a higher temperature (700K), and then anneal to the target temperature of 450K over a 2ns run. Then the system is equilibrated at 450K for 20 ns, followed by a 40 ns production run. All simulations are performed in an isothermal-isobaric (NPT) ensemble using an integration timestep of 1 fs. This procedure is similar to quench / anneal procedures widely used in IL simulations in prior works.[27,28] The short-range interactions are computed directly, and long-range electrostatic interactions are computed via the particle-particle particle mesh (PPPM) method. The pressure and temperature are controlled using the Nosé-Hoover barostat and thermostat, respectively. We perform all the simulations at 1 atm pressure. Simulations are implemented using the LAMMPS[29] open-source package. Mean square displacement (MSD) of cations and anions is calculated from equilibrium MD trajectories. Diffusivity of cation and anion in the system is calculated from the slope of the MSD plot in the normal diffusive regime. The Nernst-Einstein conductivity is calculated as[30–33] $\sigma_{NE} = \frac{1}{Vk_BT}(N_+D_+Z_+^2 + N_-D_-Z_-^2)$ where V is the volume of the system, T is the temperature, D is the diffusivity, Z is the charge, and N is the number of ions in the system. Equilibrated systems are used to perform non-equilibrium MD (NEMD) simulations to compute viscosity. In NEMD simulations, a shear is applied to the simulation box in the *xy*-plane with a constant shear rate ($\dot{\gamma} = 10^{-6}$) in the form of the SLLOD [34] equation of motion. All NEMD simulations are performed for 50 ns. The viscosity is calculated as $\eta = \frac{-\langle P_{xy}\rangle}{\dot{\gamma}}$ in the steady-steady region of the time-dependent shear stress curve. Three independent simulations are performed for each system, and all reported data are averaged over the three independent trajectories.

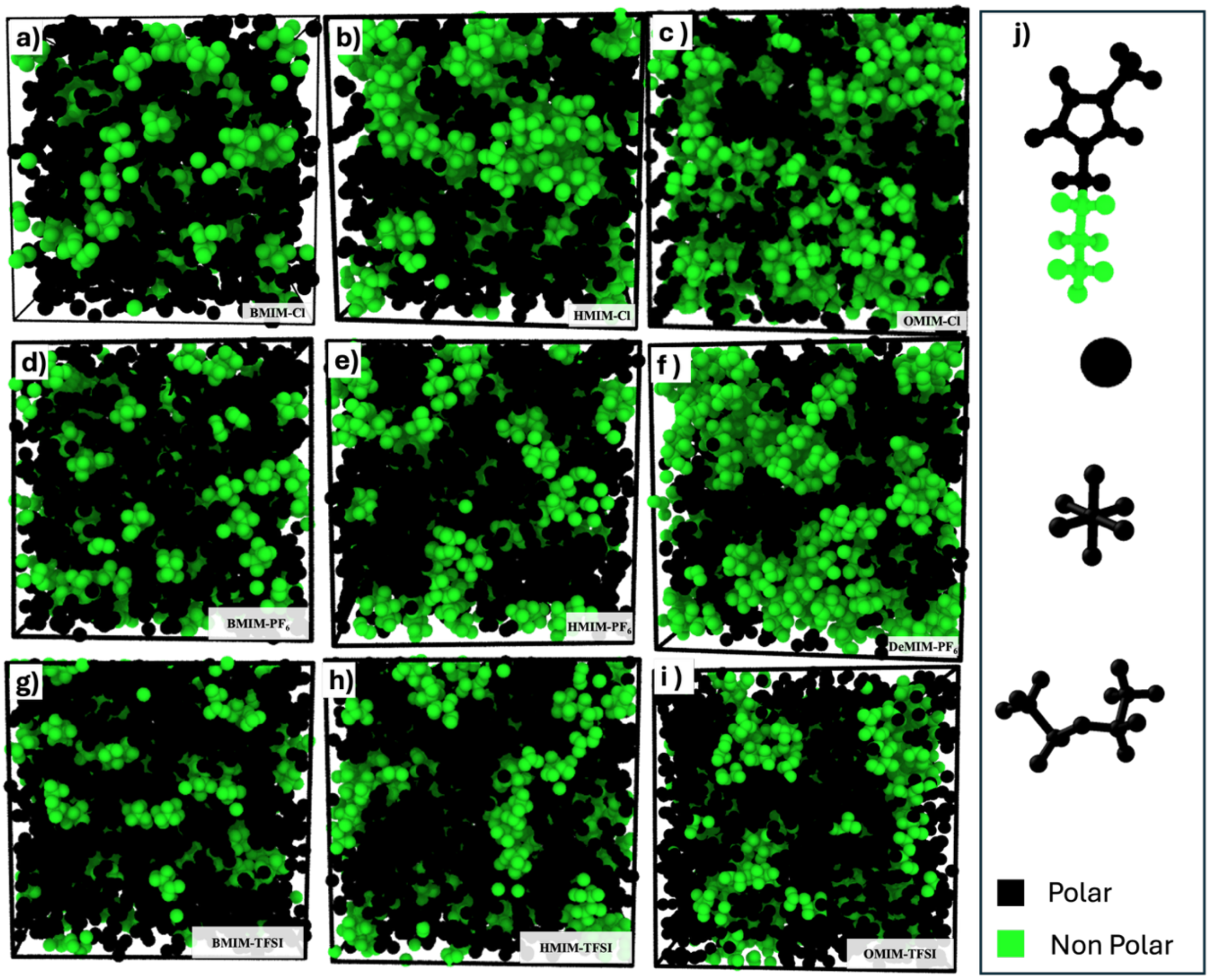

*Figure.1: (a-i) MD snapshots of 9 IL systems showing the polar and non-polar regions. (j) represents atoms correspond to polar(black) and non-polar groups (green) in a system.*

The MD snapshots of the equilibrium structures are shown in Figure 1. The polar region, which consists of dimethyl imidazolium and anions, and the non-polar region, which consists of alkyl chains, are labelled by two different colours in Figure 1. It clearly suggests that the distribution of polar and non-polar regions is strongly correlated with alkyl chain length and counterion type. We quantify this spatial variation of the polar region by analysing ion clustering. Ion clusters are identified based on their interparticle distances. Two ions are considered to be a part of a cluster when their interparticle distance is below a cut-off distance.[35–38] The cut-off distance between two chemically identical ions is taken to be the atomic diameter, whereas for chemically dissimilar ion pairs, we use the radius of the first coordination shell. For polyatomic ions, we chose a specific atom as the central atom for this cluster analysis. Phosphorus and nitrogen are chosen as the central atoms of $PF_6$ and TFSI, respectively. The carbon between two nitrogen atoms of the imidazole ring is chosen as the central atom of imidazole. This consideration is based on the previous studies,[39] which indicate that the hydrogen atom bonded to the carbon atom between the two nitrogen atoms is the most acidic site on the imidazole ring. Consequently, for the BMIM- $PF_6$ system, the highest probability distribution of $PF_6$ anion is around the top carbon of the imidazole ring. These clusters represent polar domains of a system. We estimate the volume of a polar domain by summing the atomic volumes of all the atoms that belong to it. The distribution of polar domain volume fraction is shown in Figure 2. The polar domain volume fraction ($f_{pd}$) is defined as the ratio of polar domain volume to the total system volume. Volume fraction of polar domains indicates the presence of distinct microstructural changes. Specifically, the broad distribution of polar region volume changes to a narrow distribution at the lower-volume limit as the alkyl chain length increases in the Cl ion-based systems. This leads to a reduction in the connectivity of polar domains. Moving from the Cl ion to a bigger size anion, viz. $PF_6$, we observe a bimodal distribution of polar region volume. The bimodal distribution shrinks further for TFSI. Further, for TFSI, as the alkyl chain length increases, the system shows a significant reduction in large-sized polar domains. Overall, the domain connectivity improves as the anion size increases.

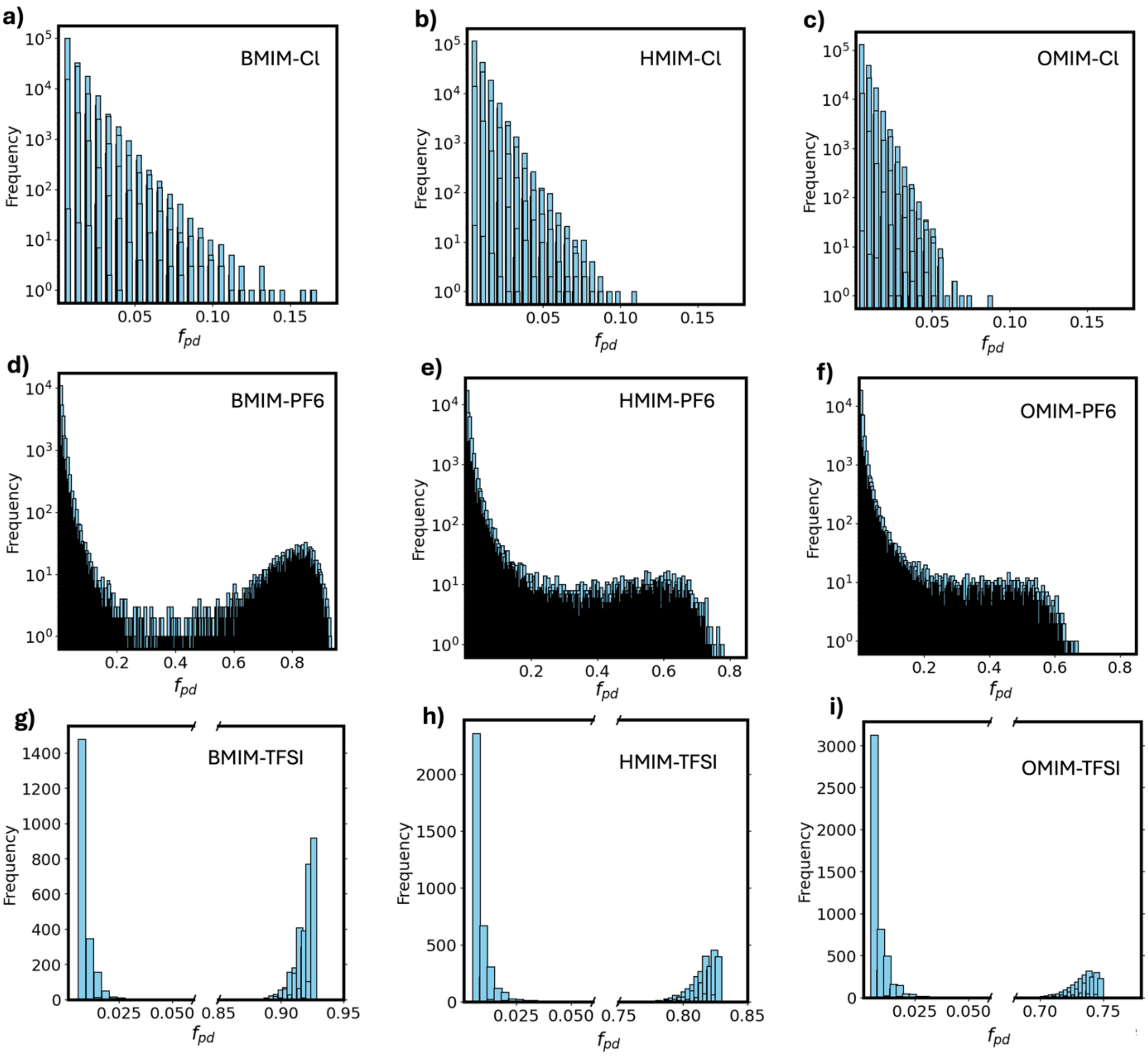

Figure 2: The polar domain volume distributions of all the systems are shown in (a-i). The polar domain volume is normalized by the system volume.

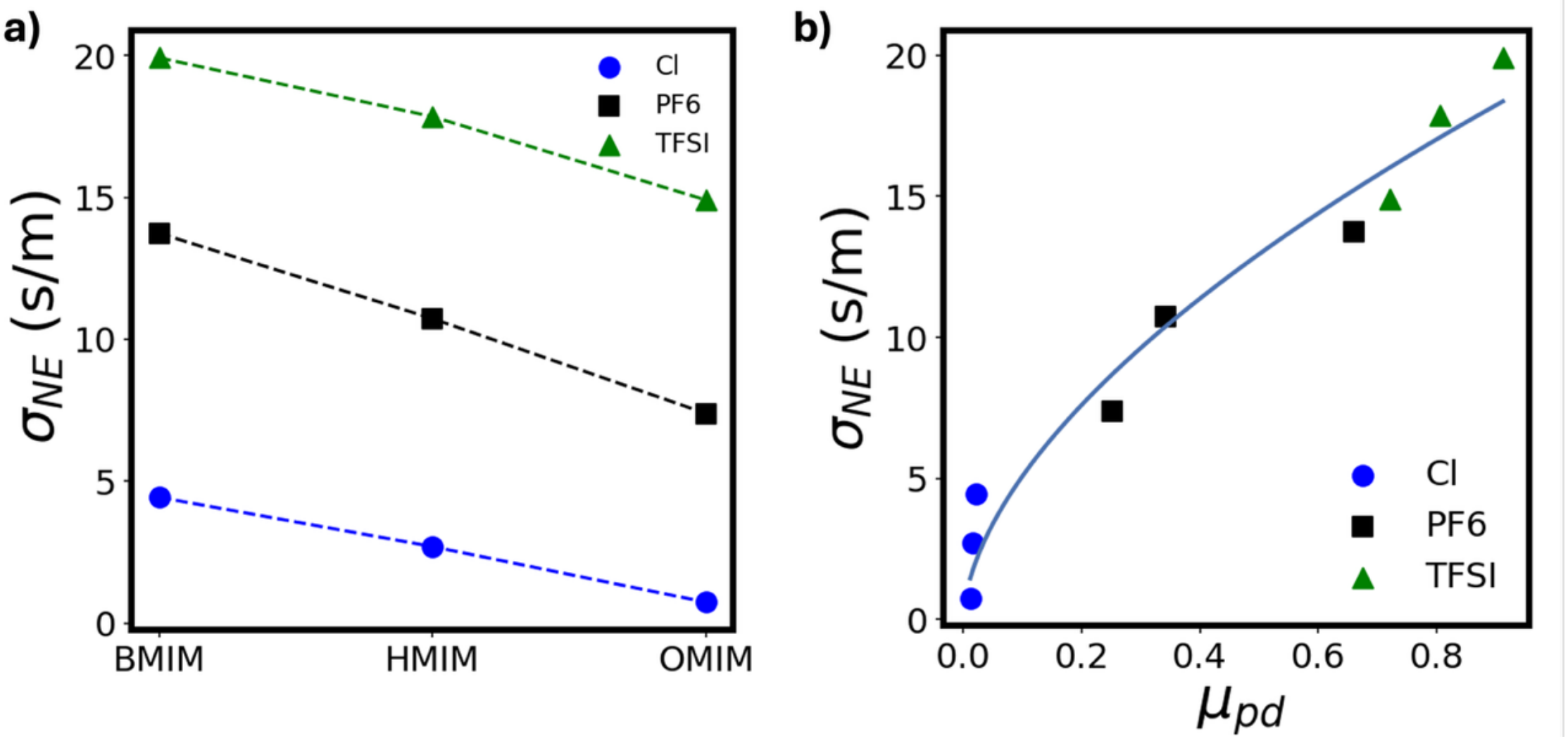

*Figure 3: Nernst-Einstein conductivity is reported in (a) for all 9 IL systems. The conductivity is plotted as a function of polar domain volume fraction in (b). The solid line in (b) corresponds to a power law of the form $\sigma_{NE} \sim \mu_{pd}^{0.58}$. The coefficient of determination ($R^2$) is 0.95.*

We now explore the connection between the polar volume distribution and the system's transport properties. Figure 3a reports the conductivity of all 9 systems. The ionic conductivity decreases monotonically with increasing alkyl chain length, indicating that longer alkyl chains progressively hinder ion transport. In addition, the conductivity exhibits a clear dependence on anion identity, following the order TFSI > $PF_6$ > Cl. The higher conductivity observed for TFSI systems can be attributed to their weaker ion-pairing and reduced electrostatic interactions, which facilitate greater ion dissociation and mobility compared with the more strongly coordinating Cl anion. Nevertheless, these three anions constitute chemically distinct classes of ILs and exhibit different correlations between ionic conductivity and alkyl-chain length. This variation raises an important question: can a common physical descriptor be identified that unifies these seemingly distinct conductivity trends across different chemical systems? We observe that the ionic conductivity across all three anion families exhibits a strong and systematic correlation with the volume of the polar domain. We find that the conductivity scales as a power law with the mean polar-domain volume fraction($\mu_{pd}$) as $\sigma_{NE} \sim \mu_{pd}^{0.58}$ (Figure 3b) with the coefficient of determination $R^2 = 0.95$. This result suggests that the polar-domain

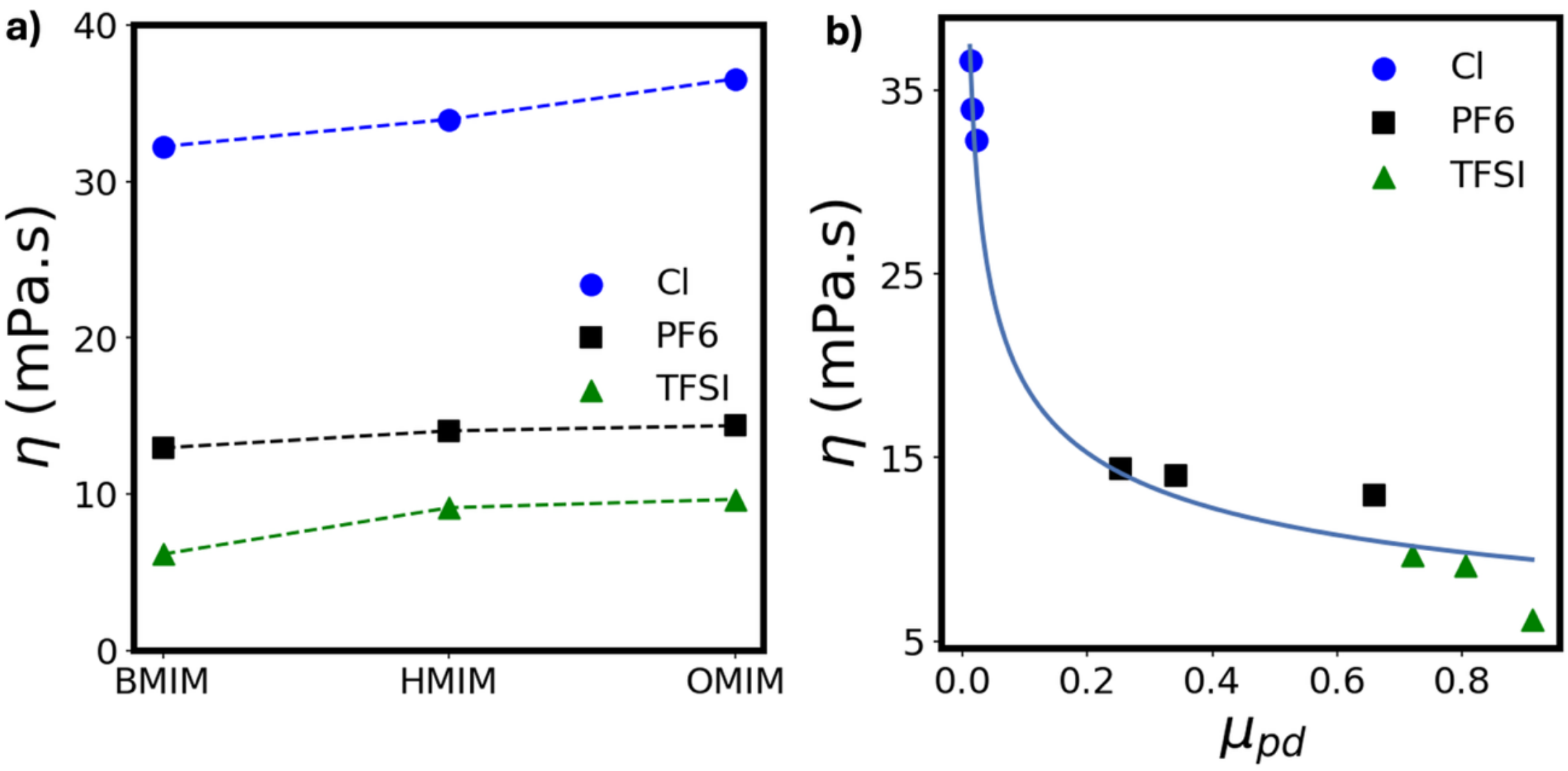

*Figure 4: The viscosities of all 9 IL systems are shown in (a) . The conductivity is plotted as a function of polar domain volume fraction in (b). The solid line in (b) corresponds to a power law of the form $\eta \sim \mu_{pd}^{-0.32}$. The coefficient of determination ($R^2$) is 0.98.*

volume provides a useful structural measure of the available ion-transport environment and captures the molecular-scale variations that arise from changes in alkyl-chain length and anion chemistry. It clearly suggests that the characteristic size of the polar domains plays a fundamental role in regulating ion transport. This power-law behavior persists across the different anion chemistries considered, indicating that the mean polar-domain volume may serve as a physically meaningful structural descriptor for the conductivity. The observed scaling further suggests that changes in molecular architecture and anion identity influence conductivity primarily through their effects on the size and organization of the polar domains, providing a potential basis for constructing a unified description of ion transport across chemically diverse ILs. Similarly, the viscosity exhibits distinct correlations with alkyl chain length for the different anion chemistries, as shown in Figure 4a. Despite these distinct chemical dependencies, the viscosity data for all systems largely collapse onto a common master curve when plotted as a function of the mean volume of the polar domains (Figure 4b). We observe that the viscosity scales as a power law with the mean polar-domain volume fraction as $\eta \sim \mu_{pd}^{-0.32}$. The coefficient of determination ($R^2$) of this power law fitting is 0.98. This collapse demonstrates that the mean polar-domain volume serves as a unifying structural descriptor that captures variations in viscosity arising from differences in anion chemistry and molecular architecture. Together with the corresponding scaling observed for ionic conductivity, these results suggest that the size of the polar domains provides a common structural basis for understanding both momentum and ion transport across chemically diverse

IL systems. We note that the polar domains exhibit highly irregular and tortuous morphologies, making it challenging to accurately quantify their volumes. We therefore approximate the volume of each polar domain as the sum of the atomic volumes of all constituent atoms. More sophisticated approaches could be employed to determine the polar domain volume, including the associated free volume. A more accurate determination of the polar domain volume may further improve the quality of the fits.

In summary, we use all-atom MD simulations to elucidate the molecular origins of transport in imidazolium-based ionic liquids and to establish quantitative structure–property relationships between their nanoscale organization and macroscopic transport behavior. By systematically varying the alkyl-chain length of the imidazolium cation and the chemical identity of the anion, we demonstrate how changes in molecular packing, ion coordination, ion aggregation, and polar–nonpolar domain organization collectively influence ionic conductivity and viscosity. Cluster and coordination analyses reveal pronounced differences in the local ionic environment across the different IL chemistries. For example, the high population of small ionic clusters in OMIM-Cl is associated with low ionic conductivity and high viscosity, whereas BMIM-TFSI exhibits a more heterogeneous distribution containing a substantial population of larger clusters, concomitant with lower viscosity and higher conductivity. These microstructural characteristics evolve systematically with both alkyl-chain length and anion identity. More importantly, we demonstrate that the seemingly distinct transport trends observed for different IL chemistries can be unified using a common structural descriptor. By characterizing the heterogeneous liquid structure through the distribution of polar-domain volumes, we find that both ionic conductivity and viscosity exhibit strong and systematic correlations with the mean polar-domain volume. As the mean polar-domain volume increases, the ionic conductivity increases whereas the viscosity decreases, with both transport properties following power-law scaling relationships. This scaling collapses the distinct anion- and chain-length-dependent trends onto common master curves, indicating that the polar-domain volume captures the essential microstructural variations governing transport across these chemically diverse systems. Our results, therefore, establish polar-domain volume as a physically motivated and efficient molecular-scale descriptor for connecting nanoscale organization with macroscopic transport properties in ILs. More broadly, the observed scaling framework provides a pathway for reducing complex molecular-level structural information to a tractable descriptor and may be applicable to other classes of heterogeneous molecular and soft matter systems in which nanoscale organization governs macroscopic properties.

## Acknowledgement

This work is made possible by financial support from the ANRF through an advanced research grant (ANRF/ARG/2025/006228/ENS). This research also uses the computational resources from Aqua and PARAM Shakti HPCE of IIT Madras. G.K.R acknowledges Prime Minister Research Fellowship (PMRF) supported by the Ministry of Education, Government of India.